\documentclass{jaa}
\usepackage{amsfonts}
\usepackage{amsmath}
\usepackage[british]{babel}
\usepackage{latexsym}
\usepackage{times}
\usepackage{balance}
\usepackage{float}
\usepackage{textcase}
\usepackage{txfonts}
\usepackage{caption}
\usepackage{color}   
\usepackage{epsfig}
\usepackage{graphics}
\usepackage{graphicx}
\usepackage{aas_macros} 
\usepackage{astrobib} 
\def\i{\item}

\newcommand{\bed}{\begin{displaymath}}
\newcommand{\eed}{\end{displaymath}}
\newcommand{\bei}{\begin{itemize}}
\newcommand{\eei}{\end{itemize}}
\newcommand{\bef}{\begin{figure}}
\newcommand{\eef}{\end{figure}}
\newcommand{\ben}{\begin{enumerate}}
\newcommand{\een}{\end{enumerate}}
\newcommand{\beq}{\begin{equation}}
\newcommand{\eeq}{\end{equation}}
\newcommand{\ber}{\begin{eqnarray}}
\newcommand{\eer}{\end{eqnarray}}

\newcommand{\pdot}{\mbox{$\dot {\rm P}$}}

\newcommand{\lsim}{\raisebox{-0.3ex}{\mbox{$\stackrel{<}{_\sim} \,$}}}

\newcounter{attnctr} 

\begin{document}\sloppy

\title{Radio Pulsar Sub-Populations (I) :  \\
  The Curious Case of Nulling Pulsars} 

\author{Sushan Konar\textsuperscript{1,*} and Uddeepta Deka\textsuperscript{1}}
\affilOne{\textsuperscript{1}NCRA-TIFR, Pune, 411007, India.\\}
\affilTwo{\textsuperscript{2}Department of Physics \& Astrophysics, University of Delhi, 110007, India.}


\twocolumn[{

\maketitle

\corres{sushan@ncra.tifr.res.in}


\begin{abstract}
About $\sim$200  radio pulsars have  been observed to  exhibit nulling
episodes -  short and long.   We find that  the nulling fraction  of a
pulsar does not have any obvious correlation with any of the intrinsic
pulsar parameters.  It also appears that the phenomenon of nulling may
be  preferentially   experienced  by  pulsars  with   emission  coming
predominantly from  the polar  cap region,  and also  having extremely
curved magnetic fields.
\end{abstract}


}]


\doinum{12.3456/s78910-011-012-3}
\artcitid{\#\#\#\#}
\volnum{000}
\year{0000}
\pgrange{1--}
\setcounter{page}{1}
\lp{1}

\section{Introduction}

Fifty  years of  observations have  yielded $\sim$3000  neutron stars,
with diverse  characteristic properties,  which fall into  three major
categories,  namely  -  a)  the rotation  powered,  b)  the  accretion
powered, and  c) the internal-energy powered  neutron stars; according
to       their       mechanisms       of       energy       generation
\cite{kaspi10,konar13,konar16c,konar17e}.  Radio pulsars, which belong
to  the  category of  rotation  powered  pulsars (RPP),  are  strongly
magnetized   rotating   neutron   stars   (mostly   isolated   or   in
non-interacting binaries)  characterized by  their short  spin periods
($P \sim 10^{-3} - 10^2$~s) and large inferred surface magnetic fields
($B  \sim 10^8  -  10^{15}$~G).   Powered by  the  loss of  rotational
energy, they emit highly coherent radiation (typically spanning almost
the  entire electromagnetic  spectrum)  which are  observed as  narrow
emission pulses.   The abrupt  cessation of  this pulsed  emission for
several pulse periods, observed in a small subset of radio pulsars, is
known as  the phenomenon of  nulling - noticed  for the first  time by
\citeN{backe70}.  Since then, close to  two hundred radio pulsars have
been observed to  experience nulling. In this context, it  needs to be
noted that most of the  $\sim$2600 radio pulsars are neither monitored
regularly, nor are  searched for the presence of  nulling.  This would
imply that two hundred  is just a lower limit to  the actual number of
nulling  pulsars. On  the other  hand,  presence of  nulling may  also
depend on the sensitivity of a  given telescope (e.g. a telescope with
a low sensitivity  may consider a pulsar to be  nulling when a similar
(or same) pulsar  might be detected in weak emission  by an instrument
with  higher sensitivity),  giving  rise to  an  over-estimate of  the
number of nulling pulsars.

In  general, two  parameters are  used to  quantify the  phenomenon of
nulling --
\vspace{-0.4cm}
\ben
   \i the  nulling fraction  (NF) - the  total percentage  fraction of
   pulses without detectable emission; and,
   \i the null length (NL) - the duration of a given nulling episode.
\een
\vspace{-0.4cm}
Both NF  and NL are observed  to span a  wide range - while  NF ranges
from just a few  to more than 90\%, NL can go from  the simple case of
single pulse nulls to the  extreme situation of complete disappearance
of  pulsed emission  for as  long as  a few  years.  Even  though most
pulsars are known to be characterised by a single value of NF (see the
tables in \ref{append} for some contrary cases), neither NF nor NL can
uniquely describe the behaviour of a  nulling pulsar. It is well known
that NL  not only  varies from  one pulsar to  another, but  also from
episode to  episode for a given  pulsar~\cite{young12}.  Moreover with
increasing  data it  is becoming  evident that  two different  pulsars
having  very different  values  of NL  and  totally different  nulling
behaviour can have the same  average value of NF~\cite{gajja12a}.  For
example,  the long  quiescent states  of intermittent  pulsars are  in
stark  contrast to  the  longest known  quiescence  times of  ordinary
nulling pulsars, i.e., they differ in their nulling timescale by about
five orders  of magnitude -  even when the  NF values are  similar for
both cases.

A detailed discussion  on different types of nulling  behaviour can be
found in  Gajjar~\citeyear{gajja17a} and references  therein.  Despite
the wide variation in NL, the population does render itself to a broad
classification, depending on the nature of nulling, as follows -
\vspace{-0.4cm}
\ben
   \i Classical  Nuller (CN) - pulsars  with mostly single (or  just a
   few)     pulse     nulls,     for     example     -     J0837-4135,
   J2022+5154~\cite{gajja12a};
   \i Intermittent  Nuller (IN) -  NL is longer,  could be up-to  a few
   hours  combined with  a longer  period of  activity, for  example -
   J1717-4054~\cite{johns92},               J1634-5107~\cite{obrie06},
   J1709-1640~\cite{naidu18};
   \i Intermittent Pulsar  (IP) - NL can vary from  days to years, for
   example   - J1933+2421~\cite{krame06a},   J1832+0029~\cite{lorim12},
   J1910+0517 \& J1929+1357~\cite{lyne17};
   \i Rotating Radio Transient (RRAT)  - Discovered in 2006, the RRATs
   are    characterised    by     their    sporadic    single    pulse
   emissions~\cite{mclau06}.   Whether these  can be  considered to  be
   part of  the nulling  fraternity is a  contentious issue, which we
   plan to take up in a later study~\cite{konar19f}.
\een

The phenomenon  of nulling is  usually observed to be  associated with
other emission features, like the {\em drifting of sub-pulses} and {\em
  mode  changing}~\cite{wang07b}.  Certain other  behavioural  changes
have also been  seen in nulling pulsars.  In  J1933+2421 the spin-down
rate has been  observed to decrease in the inactive  phase compared to
the  active phase,  suggestive of  a depletion  in the  magnetospheric
particle    outflow     in    the     quiescent    phase     of    the
pulsar~\cite{krame06,lyne09}.   An exponential  decrease in  the pulse
energy  during  a  burst  has   also  been  seen  in  certain  nulling
pulsars~\cite{ranki08,bhatt10,li12,gajja14}.   Interestingly,  nulling
behaviour   has    not   yet   been   observed    in   a   millisecond
pulsar~\cite{rajwa14}, even though the  cumulative study of this class
of pulsars is close to $10^3$ years.

In general, two different classes of models are invoked to explain the
phenomenon  of nulling,  explaining it  to arise  from -  a) intrinsic
causes  or b)  geometrical effects.   Some of  the models  attributing
nulling to an intrinsic cause are as follows -
\vspace{-0.25cm}
\bei
   \i the loss of coherence conditions~\cite{filip82};
   \i     a     complete     cessation     of     primary     particle
   production~\cite{krame06,gajja14};
   \i   changes    in   the    current   flow   conditions    in   the
   magnetosphere~\cite{timok10};
   \i a transition to a much  weaker emission mode (or an extreme case
   of mode changing)~\cite{esamd05,wang07b,timok10,young14};
   \i   time-dependent    variations   in   an    emission   `carousel
   model'~\cite{deshp01,ranki07}; etc.
\eei
\vspace{-0.25cm}
On the  other hand, a  variety of  geometrical effects have  also been
suggested to explain nulling, like -
\vspace{-0.45cm}
\bei
   \i the line-of-sight passing between emitting sub-beams giving rise
   to `pseudo-nulls'~\cite{herfi07,herfi09,ranki08};
   \i  occurrence  of various  unfavourable  changes  in the  emission
   geometry~\cite{dyks05,zhang07}.
\eei
\vspace{-0.45cm}

Detailed investigations of the nulling behaviour of individual pulsars
and theoretical modeling  of this phenomenon have  been undertaken by
many groups~\cite{ritch76,ranki86,biggs92,wang07b,gajja12a}.   In many
instances, nulling  has been  observed across  a wide  frequency range
making it a  broadband phenomenon (even though the exact  value of NF
reported appears to have large variation  over observing frequencies).
This is strongly suggestive of intrinsic changes being responsible for
nulling  rather  than  geometrical  effects.  Not  surprisingly,  many
subscribe   to  the   thought  that   nulling  is   of  magnetospheric
origin~\cite{krame06,wang07b,lyne10}.

Therefore, it is  important to look at the  overall characteristics of
the  population of  nulling pulsars  in  an effort  to understand  the
origin of the phenomenon. A  comprehensive list of nulling pulsars has
recently  been generated  by Gajjar~\citeyear{gajja17a}  comprising of
109  objects.   For  the  present  work, we  have  done  an  extensive
literature  survey to  extend and  update  that list.   The number  of
nulling  pulsars  now  stands  at   (likely  more  than)  204  (Tables
\ref{t_list01}--\ref{t_list07}).   It goes  without saying  that, like
any  such list,  this one  is incomplete.   Future observations  would
continue  to add  new nulling  pulsars to  this list,  which may  even
exhibit  hitherto unobserved  characteristic  features.  However,  the
current size of nulling pulsar population is such that it allows us to
draw certain broad conclusions about this sub-population of the larger
class of RPP. In this work, we  examine the distribution of NF and its
correlation (or  absence thereof)  with various pulsar  parameters. We
also  examine  the  general  characteristics  of  the  nulling  pulsar
population and revisit the connection of age with nulling behaviour.

\section{Characteristics of Nulling Pulsar Population}
\bef
\includegraphics[width=275pt]{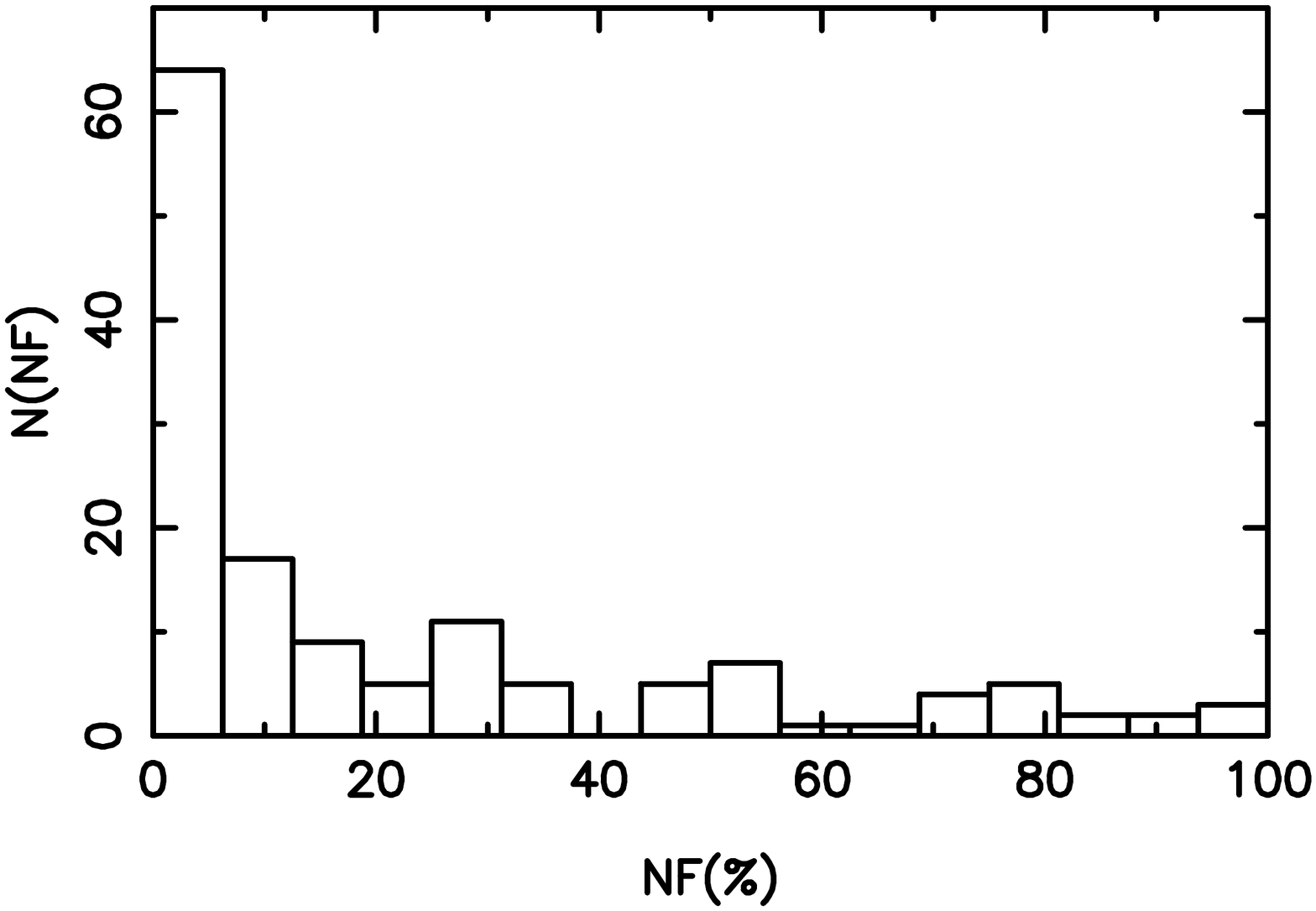}
\caption{Histogram showing  the distribution  of NF, as available  in
  the literature.  Details can be found in Tables~[2]-[5].}
\label{f_nfhst} 
\eef
Only about  8\% of all known  radio pulsars ($\sim$2500) are  known to
exhibit nulling (Tables  \ref{t_list01}--\ref{t_list07}). Quite likely
this fraction is much larger, as  only a small number of radio pulsars
are  observed  over long  periods  (or  regularly) to  detect  nulling
episodes. Also,  short nulls (nulling  episode lasting only for  a few
pulses) may not be detected in  weak pulsars.  7 among these are known
to belong to the class of Intermediate Pulsars.  Moreover, there exist
a significant number  of nulling pulsars for which no  estimate for NF
is available (Tables \ref{t_list05}--\ref{t_list06}).  Nevertheless it
is possible to draw certain broad conclusions about the population.
\begin{table} 
\begin{tabular}{|r|r|l|l|} \hline
       &&& \\
            &                  & r$_{\rm P}$ &  $\sigma$ \\
       &&& \\
NF $>$ 40\% & NF--P$_{\rm s}$    & -0.403    &  0.027 \\
            & NF--B$_{\rm s}$    & -0.220    &  0.242 \\
            & NF--\pdot        & -0.119    &  0.528 \\ 
            & NF--$\tau_{\rm c}$ & -0.047    &  0.807 \\
            & NF--DM           &  0.117    &  0.537 \\
       &&& \\
NF $<$ 40\% & NF--P$_{\rm s}$    &  0.327    &  0.0004 \\
            & NF--B$_{\rm s}$    &  0.134    &  0.160 \\
            & NF--\pdot        &  0.030    &  0.751 \\ 
            & NF--$\tau_{\rm c}$ &  0.100    &  0.295 \\
            & NF--DM           & -0.017    &  0.861 \\
       &&& \\
ALL         & NF--P$_{\rm s}$    &  0.171    &  0.044 \\
            & NF--B$_{\rm s}$    & -0.020    &  0.813 \\
            & NF--\pdot        & -0.091    &  0.283 \\
            & NF--$\tau_{\rm c}$ &  0.168    &  0.047 \\
            & NF--DM           &  0.180    &  0.033 \\
       &&& \\ \hline
\end{tabular}
\caption{Pearson's correlation  coefficient (r$_{\rm P}$) for  NF with
  various  intrinsic  pulsar  parameters and  the  significance  level
  ($\sigma$) of  the calculated value  of r$_{\rm P}$. [P$_{\rm  s}$ -
    spin-period, B$_{\rm s}$ - derived surface magnetic field, \pdot -
    spin-period derivative, $\tau_{\rm c}$  - characteristic age, DM -
    dispersion measure]}
  \label{t_corr}
\end{table}
In this context, finding a correlation  of NF with an intrinsic pulsar
parameter (spin-period, characteristic age,  magnetic field etc.)  has
been  very  important~\cite{ritch76,wang07b}.   Analysing  72  nulling
pulsars  \citeN{biggs92} found  the  spin-period (P$_{\rm  s}$) to  be
directly  proportional  to NF,  consistent  with  an earlier  work  by
\citeN{ritch76}.  Later, characteristic age ($\tau_{\rm c}$) was found
to be  correlated with  NF~\cite{wang07b}.  \citeN{corde08}  have also
reported of  finding some correlation  of the nulling  phenomenon with
small inclination angles (angle between  the rotation and the magnetic
axes).  These  observations led to  the suggestion that  older pulsars
are  harder  to  detect  as  they   spend  more  time  in  their  null
state~\cite{ritch76} and that the  phenomenon of nulling is associated
with the advanced age of a pulsar.

\begin{figure*}
\includegraphics[width=525pt]{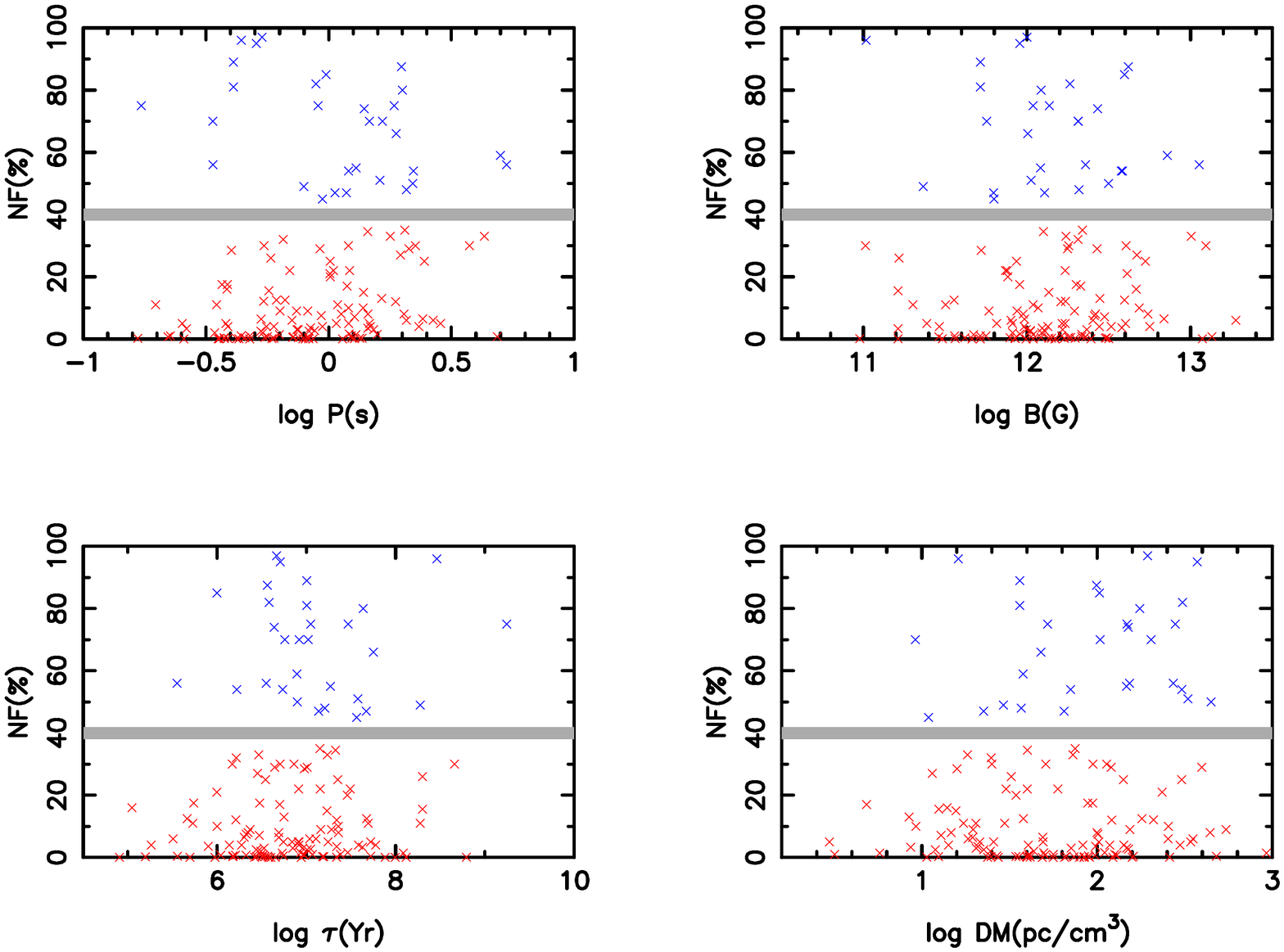}
\caption{Variation of  the nulling  fraction (NF)  against spin-period
  (P$_{\rm s}$), surface magnetic  field (B$_{\rm s}$), characteristic
  age ($\tau_{\rm c}$) and dispersion  measure (DM) of pulsars. The red
  points correspond  to pulsars with  low NF  ($<$ 40\%) and  the blue
  points to pulsars with larger values of NF. The horizontal grey band
  highlights the apparent gap in NF values around 40\%.}
\label{f_corr} 
\end{figure*}

We find,  that the NF  histogram (Fig.\ref{f_nfhst}) is  suggestive of
some kind of  bunching at lower values of NF,  and a likely separation
of NF values  at $\sim$ 40\% (although  the data size is  too small to
find  any clear  indication for  two different  NF populations).   The
general  characteristics  of  the   nulling  population,  as  seen  in
Fig.\ref{f_corr} is as follows --
\vspace{-0.25cm}
\bei
   \i $-0.5~\lsim \, \log P_s \, \lsim~0.5$ \ ;
   \i $10^{11}~\mbox{G}~\lsim \, B_s \, \lsim~10^{13}~\mbox{G}$ \ ;
   \i $10^{6}~\mbox{Yr}~\lsim \, \tau_c \, \lsim~10^{8}~\mbox{Yr}$ \ ; and
   \i $10~\mbox{pc.cm}^{-3}~\lsim \, DM \, \lsim~10^3~\mbox{pc.cm}^{-3}$ \ .
\eei
\vspace{-0.25cm}
It is evident that  there does not appear to be  any correlation of NF
with any  of the  intrinsic parameters as  per present  data.  Pearson
correlation coefficients~\cite{mises64}  calculated to find  the level
of  correlation of  NF with  various  pulsars parameters,  as seen  in
Table[\ref{t_corr}],  clearly demonstrate  this.  This  behaviour also
appears to be the same for pulsars with high as well low values of NF.

\begin{figure*} 
\includegraphics[width=525pt]{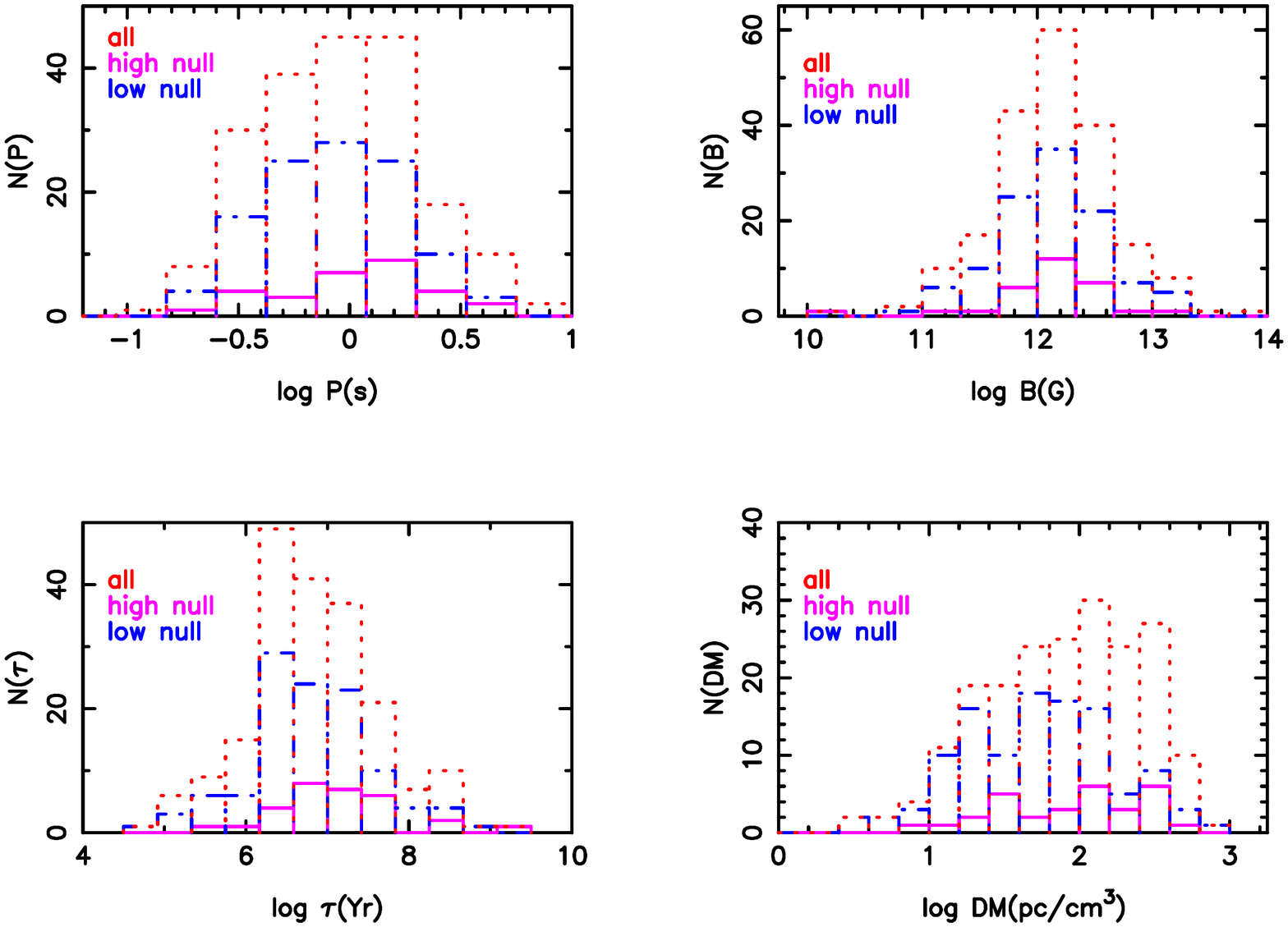}
\caption{Distribution of characteristic  pulsar parameters for pulsars
  with high null (NF  $<$ 40\%), with low null (NF  $>$ 40\%), and all
  of the  nulling pulsars.  It is  to be noted that  the histogram for
  all nulling pulsars also include pulsars without any estimate for NF
  and the intermittent pulsars.}
\label{f_nhst} 
\end{figure*}

From Fig.\ref{f_nhst},  it can  be seen  that the  earlier conjecture,
that nulling  is predominantly experienced  by old radio  pulsars with
relatively smaller  magnetic fields,  appears to be  ruled out  by the
current  population.

Interestingly,  the  nature  of  the  distribution  of  the  intrinsic
parameters appear to be very  different for pulsars exhibiting high NF
compared   to   those   having   low  values   of   NF.    A   nominal
Kolmogorov-Smirnov  (KS)  test~\cite{mises64}  on the  spin-period  of
nulling pulsars  with higher  and lower values  of NF,  yields P$_{\rm
  KS}$ = 0.002 and D$_{\rm KS}$ = 0.322, rejecting the null hypothesis
that    these   two    populations    have    the   same    underlying
distribution. [Here P$_{\rm KS}$ indicates the probability that the two
  distributions  are inherently  similar (identical),  whereas D$_{\rm
    KS}$ is the  maximum value of the absolute  difference between the
  two distributions.]  This  is also evident from  the fractional plot
of period distributions shown in Fig.\ref{f_ks}.

\bef
\vspace{-1.75cm}
\includegraphics[width=250pt]{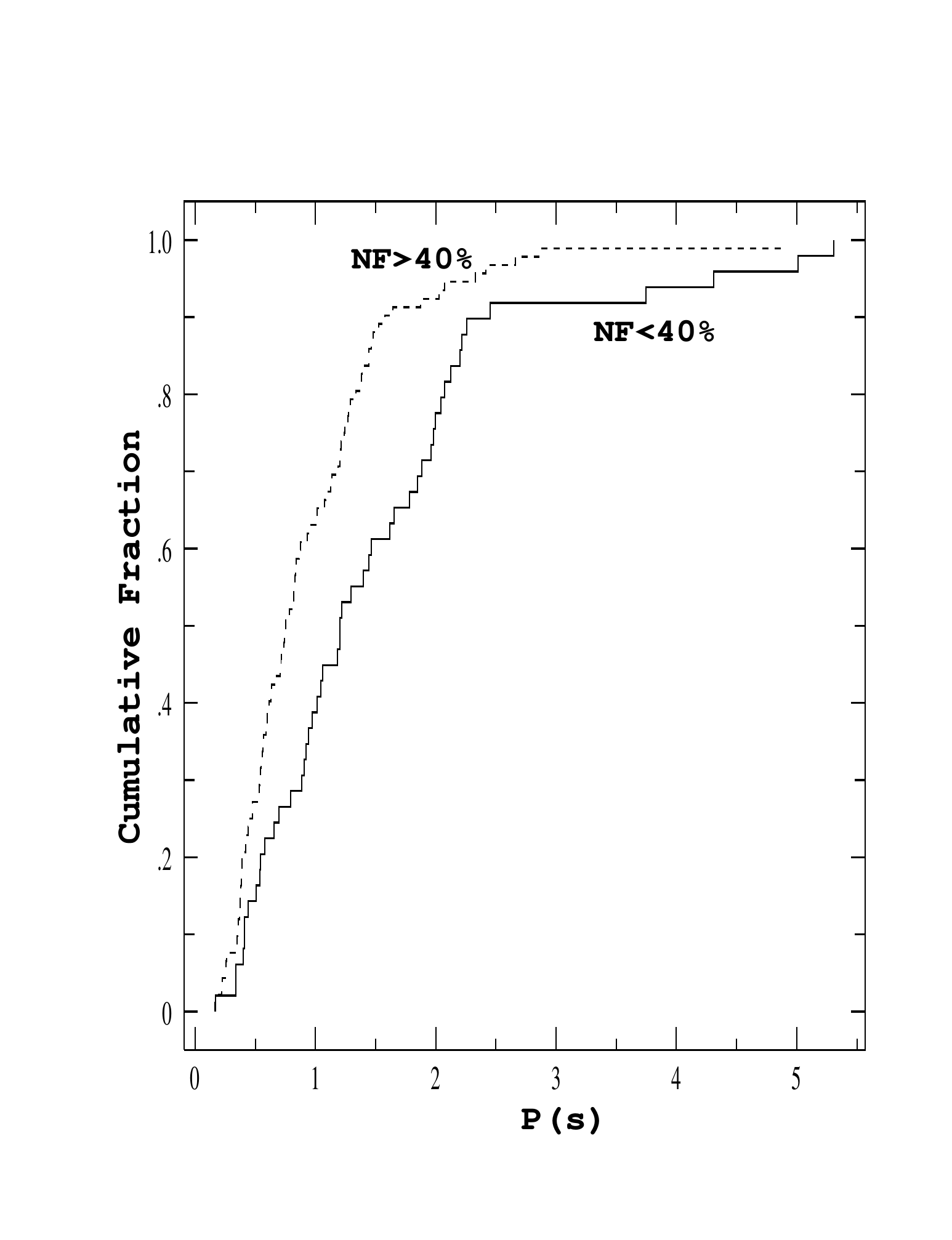}
\vspace{-0.75cm}
\caption{Cumulative fraction plot for the P$_{\rm s}$ distribution of
pulsars with low ($<$ 40\%) and high ($>$ 40\%) values of NF.}
\label{f_ks} 
\eef

\section{Pulsar Death-Lines}

Clearly, the nature  of the emission mechanism must have  a bearing on
the nulling behaviour,  whether or not nulling is  directly related to
the age of  a pulsar. As mentioned  earlier, \citeN{ritch76} undertook
the first  comprehensive study of  nulling pulsars and  suggested that
the time interval  between regular bursts of  pulse emission increases
with age, eventually leading to pulsar `death'.  This study explicitly
defined, for the very first time,  a cut-off line for pulsar emission.
This can be  thought of as the precursor of  more formal `death-line's
to  be developed  afterwards.  Later,  \citeN{zhang07} also  suggested
that nulling  pulsars are likely to  be very close to  the death line,
being active only when favourable conditions prevail.

Irrespective of  the underlying mechanism, copious  pair production in
the magnetosphere is understood to be the basic requirement for pulsar
emission. Such pair  production gives rise to a dense  plasma that can
then  allow the  growth  of  a number  of  coherent instabilities  and
generate highly  relativistic secondary  pairs which then  produce the
radio  band  emission   (see  \citeNP{mitra09},  \citeNP{melro17}  and
references therein  for details of  and recent progresses made  in the
area of pulsar emission physics).

Pulsars `switch  off' when conditions  for pair production fail  to be
met. Depending  on the  specific model, radio  pulsar `death  line' is
defined  to be  a  relation between  P$_{\rm s}$  \&  \pdot \  (period
derivative) or P$_{\rm s}$ \& B$_{\rm  s}$ beyond which the process of
pair-production  ceases  and a  pulsar  stops  emitting. A  number  of
theoretical models,  consequently a  variety of death-lines  have been
proposed to explain the present crop of pulsars.  All of these require
some  degree  of  anomalous   field  configuration  (higher  multipole
components or an offset dipole) to interpret the data in its entirety.
Some  of the  most  representative death-lines,  based upon  different
models of emission mechanism, are described below.

In the following equations - $B_{\rm p}$ is the dipolar field, $B_{\rm
  s}$ is the surface field, $r_{\rm c}$ is the radius of curvature for
the magnetic field, $h$ is the thickness  of the polar cap gap, $R$ is
the stellar  radius and  $\Omega$ is the  stellar spin  frequency. The
value of inclination angle chosen for {\bf 5b} corresponds to that for
Geminga.

{\bf A. \citeN{chen93a} :} 
  
{\bf  \em  I.  Polar Cap  Model  :  }  Pair  production ($\gamma  +  B
\rightarrow e^-  + e^+$) predominantly  happens near the polar  cap of
the neutron star~\cite{ruder75}.

{\bf 01.} Central Dipole, with $B_{\rm s} = B_{\rm p}$, $r_{\rm c} = (Rc/\Omega)^{1/2}$ -
\vspace{-0.25cm}
\beq 4 \log B - 7.5  \log P = 49.3, \label{eq_dl01} \eeq
{\bf 01a.} Dipole, off-centre by a distance $d$ -    
\vspace{-0.25cm}
\beq 4 \log B - 7.5  \log P = 49.3 - 2.5 \log [R/(R-d)], \eeq
{\bf 02.} Very curved field lines, with $r_{\rm c} \sim R$, $B_{\rm s} = B_{\rm p}$ -
\vspace{-0.25cm}
\beq 4 \log B - 6.5  \log P = 45.7, \label{eq_dl02} \eeq 
{\bf 03.} Very curved field lines, with $r_{\rm c} \sim R$,
         $B_{\rm s} = 2 \times 10^{13}$~G,
         $h \sim (B_{\rm p}/B_{\rm s})^{1/2} R (R\Omega/c)^{1/2}$ at polar cap -
\vspace{-0.25cm}
\beq 7 \log B - 13 \log P = 78, \label{eq_dl03} \eeq 
{\bf 04a, 04b.} Extremely twisted field lines, with $r_{\rm c} \sim R$ -
\vspace{-0.25cm}
\beq 4 \log B - 6  \log P = 43.8 \; \mbox{or} \; 31.3\,, \label{eq_dl04} \eeq 
(Whichever constant produces larger B in the equation above to ensure
$E_{\rm p} > 2 m_e c^2$.) 

{\bf \em II. Outer Magnetospheric  Model :} Pair production happens in
the  outer magnetosphere  via  inverse  Compton scattering,  curvature
radiation or synchrotron radiation etc.

{\bf 05a, 05b.} Aligned/Non-Aligned Dipole -
\vspace{-0.25cm}
\beq 5 \log B - 12 \log P = 72 \; \mbox{or} \; 69.5 \,. \label{eq_dl05} \eeq

{\bf B.  \citeN{zhang00a} : } In  each of the pair  of equations below
(depicted by the  sets {\bf 06a-06b, 07a-07b,  08a-08b, 09a-09b}), the
first one  corresponds to  a dipole configuration  and the  second one
corresponds to a multipolar configuration  with $B_{\rm s} \sim B_{\rm
  p}$ and  $r_{\rm c} \sim  R$.  Furthermore, $r_{\rm c6}$  is $r_{\rm
  c}$ in units of $10^6$~cm.
 
{\bf \em I. Vacuum Gap Model :} Pair production happens via formation of
a vacuum gap close to the polar cap. 

{\bf A.} Curvature Radiation -
\vspace{-0.25cm}
         \beq {\bf 06a.} \log \pdot =  11/4 \log P - 14.62, \label{eq_dl06} \eeq
         \beq {\bf 06b.} \log \pdot =  9/4  \log P - 16.58 + \log r_{\rm c6}, \label{eq_dl07} \eeq
{\bf B.} Inverse Compton Scattering -
\vspace{-0.25cm}
         \beq {\bf 07a.} \log \pdot =  2/11 \log P - 13.07, \label{eq_dl08} \eeq
         \beq {\bf 07b.} \log \pdot = -2/11 \log P - 14.50 + 8/11 \log r_{\rm c6}, \label{eq_dl09} \eeq

{\bf \em II.  Space-Charge Limited  Flow Model :} If charged particles
can be freely pulled out of  the neutron star surface, a space-charged
limited flow is generated. Mechanisms similar to those above then work
to  generate  secondary/tertiary  pairs.

{\bf A.} Curvature radiation - 
\vspace{-0.25cm}
         \beq {\bf 08a.} \log \pdot =  5/2  \log P - 14.56, \label{eq_dl10} \eeq
         \beq {\bf 08b.} \log \pdot =  2 \log P - 16.52 + \log r_{\rm c6},  \label{eq_dl11} \eeq
{\bf B.} Inverse Compton Scattering -
\vspace{-0.25cm}
         \beq {\bf 09a.} \log \pdot = -3/11 \log P - 15.36, \label{eq_dl12} \eeq
         \beq {\bf 09b.} \log \pdot = -7/11 \log P - 16.79 + 8/11 \log r_{\rm c6}. \label{eq_dl13} \eeq

\begin{figure*} 
\includegraphics[width=450pt]{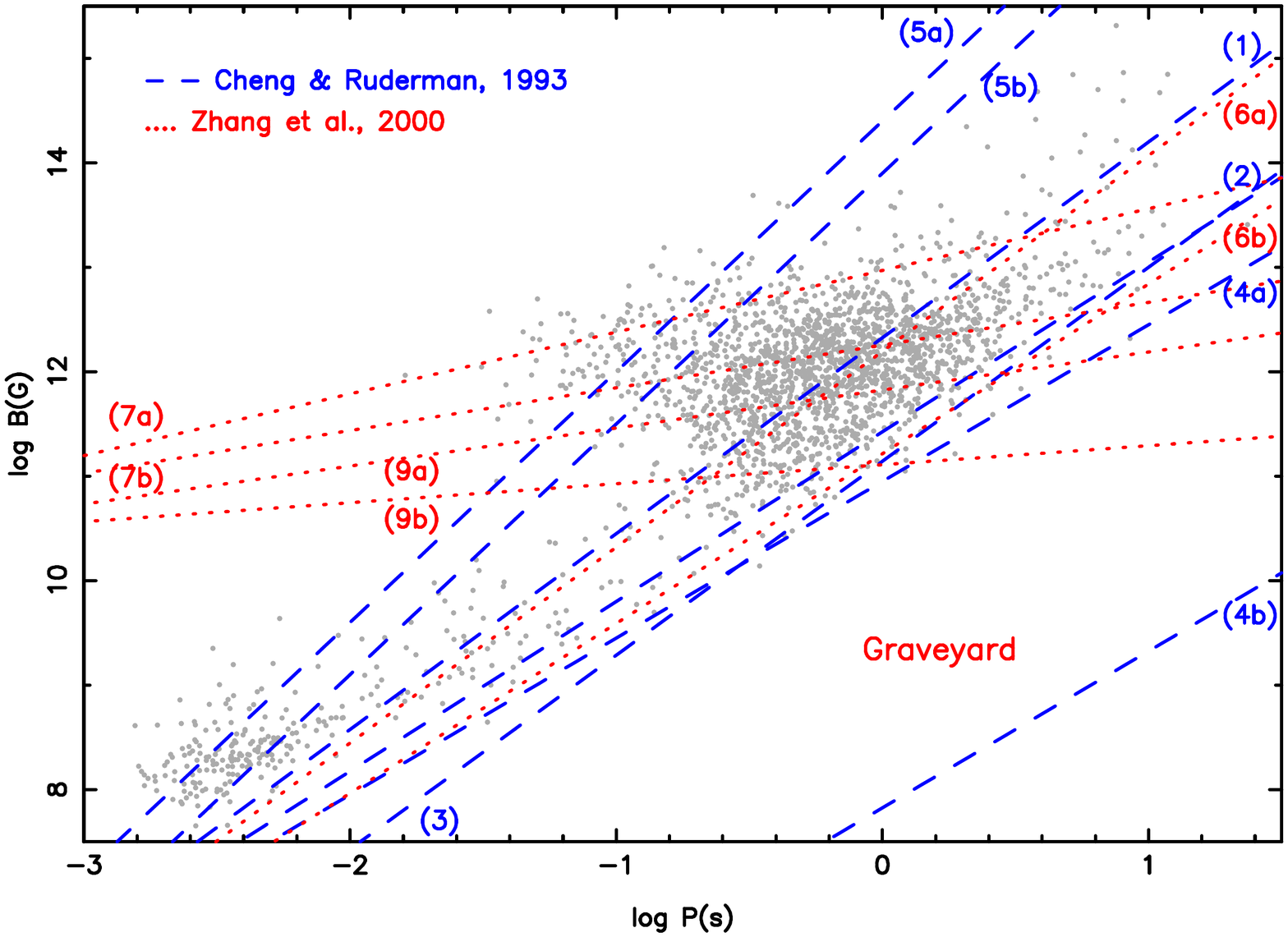}
\includegraphics[width=450pt]{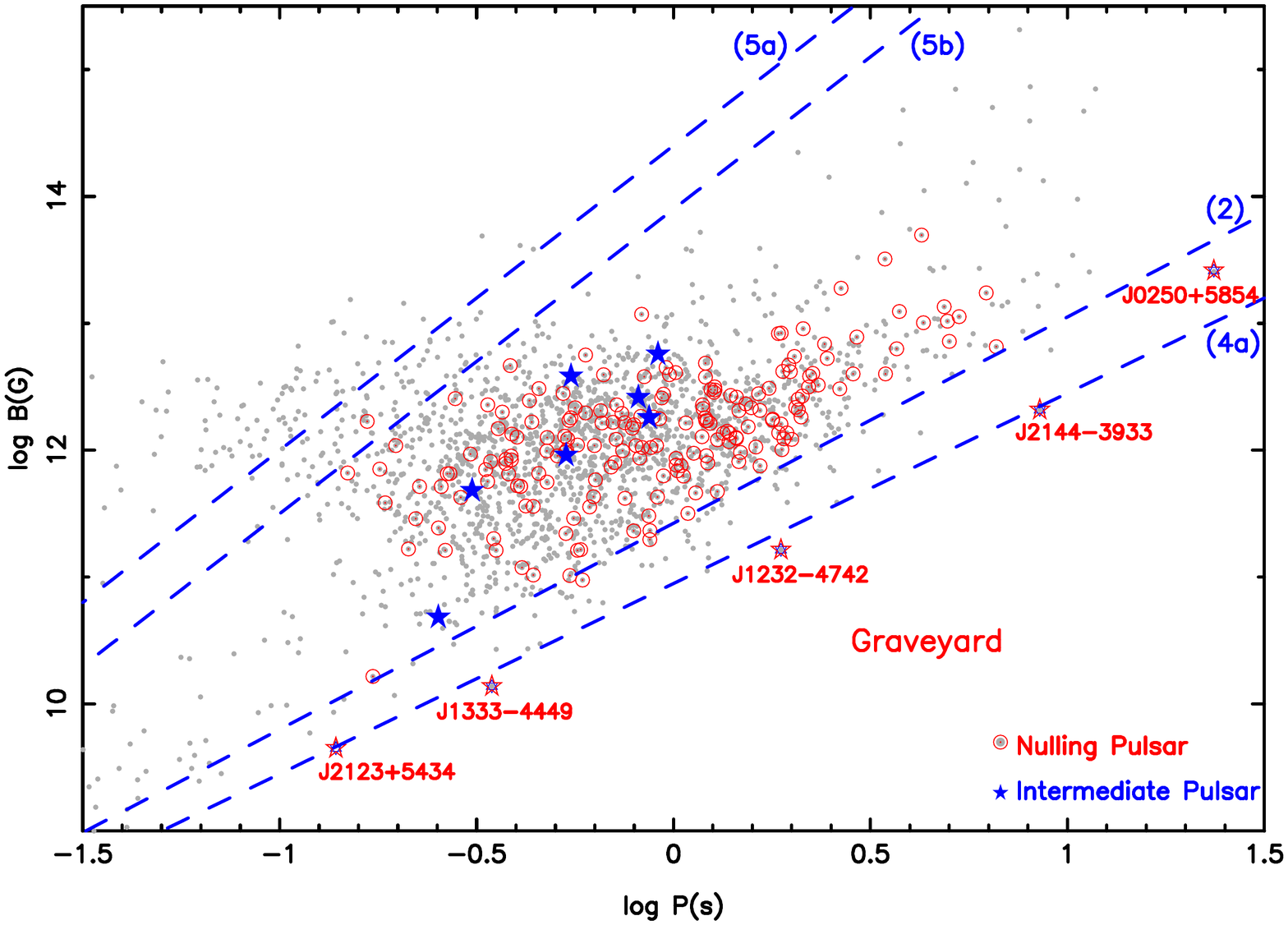}
\caption{Observed  radio pulsars  and theoretical  death-lines in  the
  P$_{\rm s}$-B$_{\rm s}$ plane. {\bf \em Top Panel} - The death lines
  have been marked according to their numbering in the text.  {\bf \em
    Bottom Panel} - Nulling pulsars and intermediate pulsars have been
  highlighted with a small subset  of death-lines. A number of pulsars
  have been  specially identified (red  open star) which appear  to be
  functioning beyond the least stringent  death line. The data for the
  known pulsars have been obtained from the ATNF pulsar catalog - {\tt
    http://www.atnf.csiro.au/research/pulsar/psrcat/}  (Manchester  et
  al., 2005). }
\label{f_death} 
\end{figure*}

All the  death-lines discussed  above have been  indicated in  the top
panel  of Fig.[\ref{f_death}],  in  the backdrop  of  the known  radio
pulsars  in P$_{\rm  s}$-B$_{\rm s}$  plane. It  should be  noted that
while the  death lines are defined  in terms of the  magnetic field by
\citeN{chen93a}, they are defined  using the derivative of spin-period
(\pdot) by  \citeN{zhang00a}.  However,  the magnetic  field is  not a
measured quantity.  An  estimate, only for the  dipolar component, is
obtained  from   the  measured   quantities  P$_{\rm  s}$   and  \pdot
\ through the following relation~\cite{manch77} -
\beq
B_{\rm p}
\simeq 3.2 \times 10^{19}
           \left(\frac{P_s}{s}\right)^{\frac{1}{2}}
           \left(\frac{\pdot}{ss^{-1}}\right)^{\frac{1}{2}}~\mbox{G}\,.
%
%
\eeq
In Fig.[\ref{f_death}], this measure of the magnetic field is used for
known pulsars. The death-line equations, given in terms of P$_{\rm s}$
and  \pdot \  by \citeN{zhang00a},  are  also plotted  in the  P$_{\rm
  s}$-B$_{\rm  s}$  plane  using  the same  measure.   Therefore,  any
conclusion drawn  for this set  of death-line equations do  not suffer
from an ambiguity regarding the measure of the magnetic field (between
the dipolar and the true surface field).  However, that is not correct
in case  of the death-lines  defined by \citeN{chen93a},  which suffer
from this  ambiguity.  It  is clear that  the death-lines  {\bf 7a-7b,
  9a-9b}  are  not  very  useful  in  constraining  the  radio  pulsar
population.   In particular,  they completely  fail to  accommodate the
millisecond pulsars. Even if one questions whether the same death-line
works for  both the  ordinary and  millisecond pulsars,  because these
equations also preclude  a significant number of  relatively low field
pulsars, we shall not consider them  hereafter. On the other hand, the
death-line depicted by  {\bf 4b} is far too deep  into the `graveyard'
to be of  much use for the current population  of pulsars. Perhaps the
newly  discovered  long-period  pulsars   (some  of  which  have  been
indicated by red stars in the bottom panel of Fig.[\ref{f_death}]) are
likely to  be constrained by this  equation.  Among the rest,  {\bf 1,
  6a} and {\bf 2, 6b} are  pairwise coincident (almost); while {\bf 2,
  3} and  {\bf 4}  envelope somewhat similar  regions.  ({\bf  8a, 8b}
are, more  or less,  coincident with  {\bf 6a,  6b} and  therefore not
shown in the figure.)

In the bottom panel of  Fig.[\ref{f_death}] the nulling pulsars (along
with intermediate pulsars)  are shown along with a  relevant subset of
death-lines.  A  number of (non-nulling)  pulsars have been  also been
identified for  their importance in  the context of  death-lines.  For
example, despite  the wide variety of  models and the large  number of
possible  death-lines  described above,  it  was  necessary to  invoke
higher multipoles, many orders of  magnitude stronger than the dipole,
at    the   the    surface   to    accommodate   the    8.5~s   pulsar
J2144-3933~\citeN{gil01}.          Other         pulsars,         like
J1232-3933~\cite{jacob09},         J1333-4449~\cite{jacob09}        or
J2123+5454~\cite{stova14} may also have  similar explanations for them
to  work beyond  the  death-line {\bf  4a}.  It is  to  be noted  that
J0250+5854~\cite{tan18}, the famous slow pulsar (P$_{\rm s}$ = 23.5s),
is  actually  within  the  allowed-zone, as  far  as  death-lines  are
concerned.

It  is  likely  that  more   than  one  emission  mechanism  could  be
responsible  for radio  pulsars activity~\cite{chen93a}.   It is  then
plausible that  different death-lines  are appropriate for  pulsars in
which different  mechanisms are responsible for  the emission. Though,
at present,  there is no  clear understanding of this.   However, when
the  population of  nulling  pulsars  are marked  out  in the  P$_{\rm
  s}$-B$_{\rm s}$  plane, certain  remarkable things are  noticed.  It
can be  seen from the  bottom panel of Fig.[\ref{f_death}]  that there
are almost no nulling pulsar above the death-line {\bf 5b} (definitely
none above  {\bf 5a}).  Now, {\bf  5a, 5b} correspond to  pure dipolar
field configurations  (aligned or non-aligned with  the rotation axis)
in an outer magnetospheric model.   Given the current understanding of
pulsar emission process, this may mean that the nulling pulsars likely
do  not possess  purely  dipolar field  configurations where  emission
originates in the outer magnetosphere.  On the other hand, the nulling
pulsars  appear to  be  bounded  below by  death-line  {\bf 2},  which
corresponds to a polar cap emission model with very curved field lines
(curvature  radius  $\sim$  stellar  radius). Taken  together,  it  is
suggestive of the  conclusion that the pulsars for  which the emission
is  predominantly  from  the  polar  cap and  the  magnetic  field  is
extremely curved are likely to experience nulling episodes.

It  has been  suggested that  in some  pulsars, the  magnetosphere may
occasionally  switch (`mode  change') between  different states  with
different    geometries    or/and     different    distributions    of
currents~\cite{timok10}. These  states have different  spin-down rates
and emission beams; and some of  the states do not (or apparently not)
have  any  radio  emission.   In  case  of  the  intermittent  pulsar,
B1931+24, it has been clearly  seen that \pdot \ significantly differs
from the off-state (nulling phase) to the on-state (active phase).  It
has been argued that, in the off-phase, the open field lines above the
magnetic pole become depleted of  charged particles and the rotational
slow down happens purely due to the magnetic dipole radiation.  On the
other  hand,  in  the  on-phase, an  additional  slow-down  torque  is
provided  by the  out-flowing  plasma~\cite{krame06a}.  Therefore,  an
estimate of the dipolar magnetic field obtained from measurements made
during the active phase is always an overestimate.  On the other hand,
\citeN{young12}  has  reported  to  have observed  no  change  in  the
spin-down rate for  the pulsar B0823+26 between the  off-state and the
on-state.  This implies  that there  would be  no overestimate  of the
dipolar field for  this pulsar.  Given this, it is  difficult to gauge
whether the reported  values of \pdot \ and hence  that of the dipolar
field  is  an  overestimate  or   not.   However,  even  with  a  10\%
overestimate (assumed for  all the nulling pulsars), we  find that our
conclusions drawn above remain unchanged.

It is clear from the  bottom-panel of Fig.[\ref{f_death}] that quite a
large number of pulsars are active  beyond the death-line {\bf 2}, but
are  bounded by  the death-line  {\bf 4a}  which again  corresponds to
polar-cap emission but the magnetic field configurations for this case
are extremely  twisted.  Because the  pulsars in this  region (between
death-lines {\bf  2} and {\bf 4a})  are slow objects with  no apparent
significance  they have  mostly not  been studied  in detail.   To our
knowledge,  there   does  not   exist  any  study   that  specifically
investigates  the  nulling  behaviour   of  pulsars  in  this  region.
However,  it  is quite  clear  that  if  these objects  are  carefully
monitored, for the presence of any  nulling episodes, we would be able
to gauge the  validity of the conclusion above. With  this in mind, we
are initiating such a program, of targeted observation of slow pulsars
between the death-lines {\bf 2} and {\bf 4a}, with the Giant Meterwave
Radio Telescope~\cite{konar19g}.

\nocite{manch05b}

\section{Summary}
About 8\%  of all known  radio pulsars  have been observed  to exhibit
nulling. In  this work, we  have considered NF  to be the  marker (for
want of  any other characteristic  parameter which has  been estimated
for a  significant number  of nulling pulsars) for a  pulsar's nulling
behaviour and have  looked at the nature of its  distribution. We have
also  considered the  general characteristics  (in terms  of intrinsic
pulsar  parameters)  of  this  sub-population of  radio  pulsars.  The
conclusions drawn are summarised as follows -
\ben
   \i There  appears to  be a  gap in the  estimated value  of nulling
   fraction  around  40\%,  separating pulsars  into  two  populations
   exhibiting higher and lower values  of NF.  However, this should be
   taken with  a bit  of caution,  as inaccurate  estimates of  NF and
   inadequate study of  pulsars with NF near 40\% could contribute
   to this bias. On the other  hand, the error bars, even though these
   could  be  quite  large   on  occasion  (tables  [\ref{t_list01}  -
     \ref{t_list04}]), do not appear likely to smudge out the gap.
   \i The number of pulsars with a lower value ($<$40\%) of NF appear to
   be far more in comparison to the  ones with a higher value of NF
   (Fig.[\ref{f_nfhst}]). Once again, this could simply be an artifact
   of observational bias.  For example, pulsars with very  high NF are
   quite likely  to be entirely  missed by rapid pulsar  surveys where
   other pulsars with  zero (or small) nulling  fractions are detected
   easily.
   \i The  distributions of  the intrinsic pulsar  parameters (P$_{\rm
     s}$,   \pdot,  B$_{\rm   s}$,  $\tau_{\rm   c}$,  DM   etc.)  are
   statistically different in these two populations of pulsars with
   high and low values of NF.
   \i There  is no evidence of any correlation of  NF with any  of the
   intrinsic pulsar parameters as per present data.  This behaviour is
   similar for pulsars with high as well as low values of NF.
   \i The  most interesting conclusion  of our study is  regarding the
   nature of the nulling pulsars.  It appears likely that pulsars, for
   which the  emission is  predominantly from the  polar cap  and have
   extremely curved magnetic fields, preferentially experience nulling
   episodes. If borne out by future observations, this would pave the
   way for a theory of nulling which has so far eluded us.
   \i Regular and targeted monitoring of pulsars in the region close to
   and bounded by the death-lines {\bf 2} and {\bf 4a} is therefore of
   great importance. As mentioned earlier, we are initiating a study with
   this goal.
\een

\section{Acknowledgment}
Most of  this work  had been carried  out when SK  was supported  by a
grant (SR/WOS-A/PM-1038/2014) from DST, Government of India and UD was
supported through  the `Indian  Academies' Summer  Research Fellowship
Programme (2017)'.   Work done by Devansh  Agarwal (also as a  part of
another  summer  project)  has  been useful  in  sorting  out  certain
preliminary issues.  SK would  also like  to thank  Avinash Deshpande,
Yashwant  Gupta, Vishal  Gajjar  and Bhal  Chandra  Joshi for  helpful
discussions. We also thank the anonymous referee for helping to improve
the clarity of the manuscript significantly.

\appendix

\section{Nulling Pulsar Parameters}
\label{append}

\begin{table*} 
\caption{Characteristic  parameters   -  spin-period   (P$_{\rm  s}$),
  surface magnetic  field (B$_{\rm s}$)  and nulling fraction  (NF) of
  known nulling  pulsars. The P$_{\rm  s}$ and B$_{\rm s}$  values are
  taken        from        the       ATNF        database        ({\tt
    http://www.atnf.csiro.au/research/pulsar/psrcat/}) while NF values
  have been indicated with  appropriate references, inclusive of cases
  where different estimates have been reported by different groups.}
  \label{t_list01}
\begin{minipage}{0.95\textwidth}
\begin{tabular}{|r l l l l l l|} \hline
       &&&&&& \\
       &  PR'S Name      &    J-Name        &   P$_{\rm s}$     &    B$_{\rm s}$              &   NF               & References        \\
       &                &                  &   (s)           &    (G)                     &   (\%)           &                   \\ 
       &&&&&& \\
   1   &  B0031-07      &  J0034-0721      &   0.9429        &    $6.28 \times 10^{11}$   &   $44.0\pm1.0$   &   \citeNP{gajja17a} \\
   2   &  B0045+33      &  J0048+3412      &   1.2171        &    $1.71 \times 10^{12}$   &   $21.0\pm1.0$   &   \citeNP{redma09}  \\
   3   &  B0148-06      &  J0151-0635      &   1.4647        &    $8.15 \times 10^{11}$   &   $\le5.0$       &   \citeNP{biggs92}  \\
   4   &  B0149-16      &  J0152-1637      &   0.8327        &    $1.05 \times 10^{12}$   &   $\le2.5$       &   \citeNP{vivek95}  \\
   5   &  B0301+19      &  J0304+1932      &   1.3876        &    $1.36 \times 10^{12}$   &   $10.0$         &   \citeNP{ranki86}  \\
   6   &  B0329+54      &  J0332+5434      &   0.7145        &    $1.22 \times 10^{12}$   &   $\le0.25$      &   \citeNP{ritch76}  \\
   7   &  B0450-18      &  J0452-1759      &   0.5489        &    $1.80 \times 10^{12}$   &   $\le0.5$       &   \citeNP{ritch76}  \\
   8   &  J0458-0505    &  J0458-0505      &   1.8835        &    $1.01 \times 10^{12}$   &   $63.0\pm3.0$   &   \citeNP{lynch13}  \\
   9   &  B0523+11      &  J0525+1115      &   0.3544        &    $1.63 \times 10^{11}$   &   $\le0.06$      &   \citeNP{weisb86}  \\
  10   &  B0525+21      &  J0528+2200      &   3.7455        &    $1.24 \times 10^{13}$   &   $25.0\pm5.0$   &   \citeNP{ritch76}  \\
  11   &  B0529-66      &  J0529-6652      &   0.9757        &    $3.94 \times 10^{12}$   &   $83.5\pm1.5$   &   \citeNP{crawf13}  \\
  12   &  B0626+24      &  J0629+2415      &   0.4766        &    $9.87 \times 10^{11}$   &   $\le0.02$      &   \citeNP{weisb86}  \\
  13   &  B0628-28      &  J0630-2834      &   1.2444        &    $3.01 \times 10^{12}$   &   $\le0.3$       &   \citeNP{biggs92}  \\
  14   &  B0656+14      &  J0659+1414      &   0.3849        &    $4.66 \times 10^{12}$   &   $12.0\pm4.0$   &   \citeNP{weisb86}  \\
  15   &  B0736-40      &  J0738-4042      &   0.3749        &    $7.88 \times 10^{11}$   &   $\le0.4$       &   \citeNP{biggs92}  \\
  16   &  B0740-28      &  J0742-2822      &   0.1668        &    $1.69 \times 10^{12}$   &   $\le0.2$       &   \citeNP{biggs92}  \\
  17   &  B0751+32      &  J0754+3231      &   1.4423        &    $1.26 \times 10^{12}$   &   $34.0\pm0.5$   &   \citeNP{weisb86}  \\
  18   &  B0809+74      &  J0814+7429      &   1.2922        &    $4.72 \times 10^{11}$   &   $\le 5.0$      &   \citeNP{ritch76}  \\
  19   &  B0818-13      &  J0820-1350      &   1.2381        &    $1.63 \times 10^{12}$   &   $1.5\pm0.25$   &   \citeNP{ritch76}  \\
  20   &  B0818-41      &  J0820-4114      &   0.5454        &    $1.03 \times 10^{11}$   &   $30.0$         &   \citeNP{bhatt10}   \\
  21   &  B0820+02      &  J0823+0159      &   0.8649        &    $3.04 \times 10^{11}$   &   $\le0.06$      &   \citeNP{weisb86}  \\
  22   &  B0823+26      &  J0826+2637      &   0.5307        &    $9.64 \times 10^{11}$   &   $6.4\pm0.8$    &  \citeNP{ranki95}   \\
  23   &  B0826-34      &  J0828-3417      &   1.8489        &    $1.37 \times 10^{12}$   &   $75.0\pm35.0$  &   \citeNP{durdi79}  \\
  24   &  B0833-45      &  J0835-4510      &   0.0893        &    $3.38 \times 10^{12}$   &   $\le0.0008$    &   \citeNP{biggs92}  \\
  25   &  B0834+06      &  J0837+0610      &   1.2738        &    $2.98 \times 10^{12}$   &   $7.1\pm0.1$    &   \citeNP{ritch76}  \\
  26   &  B0835-41      &  J0837-4135      &   0.7516        &    $1.65 \times 10^{12}$   &   $1.7\pm1.2$    &   \citeNP{gajja17a} \\
  27   &  B0906-17      &  J0908-1739      &   0.4016        &    $5.25 \times 10^{11}$   &   $26.8\pm1.7$   &   \citeNP{basu17}   \\
       &                &                  &                 &    $5.25 \times 10^{11}$   &   $25.7\pm1.3$   &   \citeNP{basu17}   \\
  28   &  B0919+06      &  J0922+0638      &   0.4306        &    $2.46 \times 10^{12}$   &   $\le0.05$      &   \citeNP{weisb86}  \\
  29   &  B0932-52      &  J0934-5249      &   1.4448        &    $2.62 \times 10^{12}$   &   $5.0\pm3.0$    &   \citeNP{naidu17}  \\
  30   &  B0940-55      &  J0942-5552      &   0.6644        &    $3.94 \times 10^{12}$   &   $\le12.5$      &   \citeNP{biggs92}  \\
  31   &  B0940+16      &  J0943+1631      &   1.0874        &    $3.18 \times 10^{11}$   &   $8.0\pm3.0$    &   \citeNP{weisb86}  \\
  32   &  B0942-13      &  J0944-1354      &   0.5703        &    $1.63 \times 10^{11}$   &   $14.4\pm0.9$   &   \citeNP{basu17}   \\
       &                &                  &                 &                           &   $\le7.0$       &   \citeNP{vivek95}  \\
  33   &  B0950+08      &  J0953+0755      &   0.2531        &    $2.44 \times 10^{11}$   &   $\le5.0$       &   \citeNP{ritch76}  \\
  34   &  J1049-5833    &  J1049-5833      &   2.2023        &    $3.15 \times 10^{12}$   &   $47.0\pm3.0$   &   \citeNP{wang07b}  \\
       &                &                  &                 &                           &   $47.0\pm3.0$   &   \citeNP{yang14}   \\ 
  35   &  B1055-52      &  J1057-5226      &   0.1971        &    $1.09 \times 10^{12}$   &   $\le11.0$      &   \citeNP{biggs92}  \\
  36   &  B1112+50      &  J1115+5030      &   1.6564        &    $2.06 \times 10^{12}$   &   $64.0\pm6.0$   &   \citeNP{gajja17a} \\
  37   &  B1114-41      &  J1116-4122      &   0.9432        &    $2.77 \times 10^{12}$   &   $3.3\pm0.5$    &   \citeNP{basu17}   \\
  38   &  B1133+16      &  J1136+1551      &   1.1879        &    $2.13 \times 10^{12}$   &   $15.0\pm2.5$   &   \citeNP{ritch76}  \\
  39   &  B1237+25      &  J1239+2453      &   1.3824        &    $1.17 \times 10^{12}$   &   $6.0\pm2.5$    &   \citeNP{ritch76}  \\
       &                &                  &                 &                           &   $7.0\pm3.0$    &   \citeNP{naidu17}  \\
  40   &  B1240-64      &  J1243-6423      &   0.3885        &    $1.34 \times 10^{12}$   &   $\le4.0$       &   \citeNP{biggs92}  \\
       &&&&&& \\ \hline
\end{tabular}
\end{minipage}
\end{table*}
\begin{table*} 
  \caption{Continuation of Table~\ref{t_list01}.}
  \label{t_list02}
\begin{minipage}{0.95\textwidth}
\begin{tabular}{|r l l l l l l|} \hline
       &&&&&& \\
       &  PSR Name      &    J-Name        &   P$_{\rm s}$     &    B$_{\rm s}$              &   NF               & References        \\
       &                &                  &   (s)           &    (G)                     &   (\%)           &                   \\ 
       &&&&&& \\
  41   &  B1322-66      &  J1326-6700      &   0.5430        &    $1.72 \times 10^{12}$   &   $9.1\pm3.0$    &   \citeNP{wang07b}  \\
  42   &  B1325-49      &  J1328-4921      &   1.4787        &    $9.61 \times 10^{11}$   &   $4.0$          &   \citeNP{basu17}   \\
  43   &  B1358-63      &  J1401-6357      &   0.8428        &    $3.80 \times 10^{12}$   &   $1.6\pm2.0$    &   \citeNP{wang07b}  \\
  44   &  B1426-66      &  J1430-6623      &   0.7854        &    $1.49 \times 10^{12}$   &   $\le0.05$      &   \citeNP{biggs92}  \\
  45   &  B1451-68      &  J1456-6843      &   0.2634        &    $1.63 \times 10^{11}$   &   $\le3.3$       &   \citeNP{biggs92}  \\
  46   &  J1502-5653    &  J1502-5653      &   0.5355        &    $9.99 \times 10^{11}$   &   $93.0\pm4.0$   &   \citeNP{wang07b}  \\
  47   &  B1508+55      &  J1509+5531      &   0.7397        &    $1.95 \times 10^{12}$   &   $7.0\pm2.0$    &   \citeNP{naidu17}  \\
  48   &  J1525-5417    &  J1525-5417      &   1.0117        &    $4.09 \times 10^{12}$   &   $16.0\pm5.0$   &   \citeNP{wang07b}  \\
  49   &  B1524-39      &  J1527-3931      &   2.4176        &    $6.87 \times 10^{12}$   &   $5.1\pm1.3$    &   \citeNP{basu17}   \\
  50   &  B1530+27      &  J1532+2745      &   1.1248        &    $9.48 \times 10^{11}$   &   $6.0\pm2.0$    &   \citeNP{weisb86}  \\
  51   &  B1530-53      &  J1534-5334      &   1.3689        &    $1.41 \times 10^{12}$   &   $\le0.25$      &   \citeNP{biggs92}  \\
  52   &  B1540-06      &  J1543-0620      &   0.7091        &    $7.99 \times 10^{11}$   &   $4.0\pm2.0$    &   \citeNP{naidu17}  \\
  53   &  B1556-44      &  J1559-4438      &   0.2571        &    $5.18 \times 10^{11}$   &   $\le0.01$      &   \citeNP{biggs92}  \\
       &                &                  &                 &                           &   $0.24$         &   \citeNP{basu17}   \\
  54   &  B1604-00      &  J1607-0032      &   0.4218        &    $3.64 \times 10^{11}$   &   $\le0.1$       &   \citeNP{biggs92}  \\
  55   &  B1612+07      &  J1614+0737      &   1.2068        &    $1.71 \times 10^{12}$   &   $\le5.0$       &   \citeNP{weisb86}  \\ 
  56   &  J1634-5107    &  J1634-5107      &   0.5074        &    $9.04 \times 10^{11}$   &   $90.0\pm5.0$   &   \citeNP{young15}  \\
  57   &  J1639-4359    &  J1639-4359      &   0.5876        &    $9.50 \times 10^{10}$   &   $\le0.1$       &   \citeNP{gajja17a} \\
  58   &  B1641-45      &  J1644-4559      &   0.4551        &    $3.06 \times 10^{12}$   &   $\le0.4$       &   \citeNP{biggs92}  \\
  59   &  B1642-03      &  J1645-0317      &   0.3877        &    $8.41 \times 10^{11}$   &   $\le0.25$      &   \citeNP{ritch76}  \\
  60   &  J1648-4458    &  J1648-4458      &   0.6296        &    $1.09 \times 10^{12}$   &   $1.4$          &   \citeNP{wang07b}  \\
  61   &  J1649+2533    &  J1649+2533      &   1.0153        &    $7.63 \times 10^{11}$   &   $\le20.0$      &   \citeNP{redma09}  \\
  62   &  B1658-37      &  J1701-3726      &   2.4546        &    $5.29 \times 10^{12}$   &   $14.0\pm2.0$   &   \citeNP{yang14}   \\
       &                &                  &                 &                           &   $19.0\pm6.0$   &   \citeNP{gajja17a} \\
  63   &  J1702-4428    &  J1702-4428      &   2.1235        &    $2.68 \times 10^{12}$   &   $26.0\pm3.0$   &   \citeNP{wang07b}  \\
  64   &  B1700-32      &  J1703-3241      &   1.2118        &    $9.05 \times 10^{11}$   &   $1.6\pm0.4$    &   \citeNP{basu17}   \\
  65   &  J1703-4851    &  J1703-4851      &   1.3964        &    $2.70 \times 10^{12}$   &   $1.1$          &   \citeNP{wang07b}  \\
       &                &                  &                 &                           &   $74.0$         &   \citeNP{yang14}   \\
  66   &  B1706-16      &  J1709-1640      &   0.6531        &    $2.05 \times 10^{12}$   & $31.0\pm2.0$, $15.0$ & \citeNP{naidu18} \\
  67   &  J1715-4034    &  J1715-4034      &   2.0722        &    $2.53 \times 10^{12}$   &   $\le6.0$       &   \citeNP{gajja17a} \\
  68   &  B1713-40      &  J1717-4054      &   0.8877        &    $1.83 \times 10^{12}$   &   $77.0\pm5.0$   &   \citeNP{young15}  \\ 
       &                &                  &                 &                           &   $\ge 95.0$     &   \citeNP{wang07b}  \\ 
  69   &  B1718-32      &  J1722-3207      &   0.4772        &    $5.62 \times 10^{11}$   &   $1.0\pm1.0$    &   \citeNP{naidu17}  \\
  70   &  J1725-4043    &  J1725-4043      &   1.4651        &    $2.05 \times 10^{12}$   &   $\le70.0$      &   \citeNP{gajja17a} \\
  71   &  J1727-2739    &  J1727-2739      &   1.2931        &    $1.21 \times 10^{12}$   &   $52.0\pm3.0$   &   \citeNP{wang07b}  \\
  72   &  B1727-47      &  J1731-4744      &   0.8298        &    $1.18 \times 10^{13}$   &   $\le0.1$       &   \citeNP{biggs92}  \\
  73   &  B1730-37      &  J1733-3716      &   0.3376        &    $2.28 \times 10^{12}$   &   $52.4\pm3.5$   &   \citeNP{basu17}   \\
  74   &  J1738-2330    &  J1738-2330      &   1.9788        &    $4.16 \times 10^{12}$   &   $85.1\pm2.3$   &   \citeNP{gajja17a} \\
  75   &  B1737+13      &  J1740+1311      &   0.8031        &    $1.09 \times 10^{12}$   &   $\le0.02$      &   \citeNP{weisb86}  \\
  76   &  B1738-08      &  J1741-0840      &   2.0431        &    $2.18 \times 10^{12}$   &   $30.0\pm5.0$   &   \citeNP{gajja17b} \\
       &                &                  &                 &                           & $15.7\pm1.7$, $15.8\pm1.4$ & \citeNP{basu17} \\
  77   &  J1744-3922    &  J1744-3922      &   0.1724        &    $1.65 \times 10^{10}$   &   $\le75.0$      &   \citeNP{faulk04}  \\
  78   &  B1742-30      &  J1745-3040      &   0.3674        &    $1.99 \times 10^{12}$   &   $\le17.5$      &   \citeNP{biggs92}  \\
  79   &  B1747-46      &  J1751-4657      &   0.7424        &    $9.91 \times 10^{11}$   &   $2.4\pm0.5$    &   \citeNP{basu17}   \\
  80   &  J1752+2359    &  J1752+2359      &   0.4091        &    $5.19 \times 10^{11}$   &   $\le89.0$      &   \citeNP{gajja17a} \\
  81   &  B1749-28      &  J1752-2806      &   0.5626        &    $2.16 \times 10^{12}$   &   $\le0.75$      &   \citeNP{ritch76}  \\
  82   &  J1752+2359    &  J1752+2359      &   0.4091        &    $5.19 \times 10^{11}$   &   $81.0$         &   \citeNP{yang14}   \\ 
  83   &  B1758-03      &  J1801-0357      &   0.9215        &    $1.77 \times 10^{12}$   & $27.7\pm1.3$, $26.1\pm2.6$ & \citeNP{basu17} \\
  84   &  J1808-0813    &  J1808-0813      &   0.8760        &    $1.05 \times 10^{12}$   &   $1.28\pm1.3$   &   \citeNP{basu17}   \\
  85   &  B1809-173     &  J1812-1718      &   1.2054        &    $4.85 \times 10^{12}$   &   $5.8\pm0.4$    &   \citeNP{wang07b}  \\ 
       &&&&&& \\ \hline
\end{tabular}
\end{minipage}
\end{table*}
\begin{table*} 
  \caption{Continuation of Table~\ref{t_list01} \& \ref{t_list02}.}
  \label{t_list03}
\begin{minipage}{0.95\textwidth}
\begin{tabular}{|r l l l l l l|} \hline
       &&&&&& \\
       &  PSR Name      &    J-Name        &   P$_{\rm s}$     &    B$_{\rm s}$              &   NF               & References        \\
       &                &                  &   (s)           &    (G)                     &   (\%)           &                   \\ 
       &&&&&& \\
  86   &  B1813-36      &  J1817-3618      &   0.3870        &    $9.01 \times 10^{11}$   &   $16.7\pm0.7$   &   \citeNP{basu17}   \\
  87   &  J1819+1305    &  J1819+1305      &   1.0604        &    $6.25 \times 10^{11}$   &   $41.0\pm6.0$   &   \citeNP{yang14}   \\
  88   &  B1818-04      &  J1820-0427      &   0.5981        &    $1.97 \times 10^{12}$   &   $\le0.25$      &   \citeNP{biggs92}  \\
  89   &  J1820-0509    &  J1820-0509      &   0.3373        &    $5.67 \times 10^{11}$   &   $67.0\pm3.0$   &   \citeNP{wang07b}  \\
  90   &  B1819-22      &  J1822-2256      &   1.8743        &    $1.61 \times 10^{12}$   &   $10.0\pm2.0$   &   \citeNP{naidu17}  \\
       &                &                  &                 &                           &   $4.7\pm0.9$    &   \citeNP{basu17} \\
       &                &                  &                 &                           &   $5.5\pm0.7$    &   --do--            \\
  91   &  B1821+05      &  J1823+0550      &   0.7529        &    $4.18 \times 10^{11}$   &   $\le0.4$       &   \citeNP{weisb86}  \\
  92   &  J1831-1223    &  J1831-1223      &   2.8580        &    $3.99 \times 10^{12}$   &   $4.0\pm1.0$    &   \citeNP{wang07b}  \\
  93   &  J1833-1055    &  J1833-1055      &   0.6336        &    $5.85 \times 10^{11}$   &   $7.0\pm2.0$    &   \citeNP{wang07b}  \\
  94   &  J1840-0840    &  J1840-0840      &   5.3094        &    $1.13 \times 10^{13}$   &   $50.0\pm6.0$   &   \citeNP{gajja17b} \\
  95   &  B1839+09      &  J1841+0912      &   0.3813        &    $6.52 \times 10^{11}$   &   $\le5.0$       &   \citeNP{weisb86}  \\
  96   &  J1843-0211    &  J1843-0211      &   2.0275        &    $5.48 \times 10^{12}$   &   $6.0\pm2.0$    &   \citeNP{wang07b}  \\
  97   &  B1842+14      &  J1844+1454      &   0.3755        &    $8.48 \times 10^{11}$   &   $\le0.15$      &   \citeNP{weisb86}  \\ 
  98   &  B1844-04      &  J1847-0402      &   0.5978        &    $5.63 \times 10^{12}$   &   $3.0\pm1.0$    &   \citeNP{naidu17}  \\
  99   &  B1845-19      &  J1848-1952      &   4.3082        &    $1.01 \times 10^{13}$   &   $27.0\pm6.0$   &   \citeNP{naidu17}  \\
 100   &  B1848+12      &  J1851+1259      &   1.2053        &    $3.77 \times 10^{12}$   &   $\le54.0$      &   \citeNP{redma09}  \\
 101   &  J1853+0505    &  J1853+0505      &   0.9051        &    $1.09 \times 10^{12}$   &   $67.0\pm8.0$   &   \citeNP{young15}  \\
 102   &  B1857-26      &  J1900-2600      &   0.6122        &    $3.58 \times 10^{11}$   &   $10.0\pm2.5$   &   \citeNP{ritch76}  \\
 103   &  J1901+0413    &  J1901+0413      &   2.6631        &    $1.89 \times 10^{13}$   &   $\le6.0$       &   \citeNP{gajja17a} \\
 104   &  J1901-0906    &  J1901-0906      &   1.7819        &    $1.73 \times 10^{12}$   &   $29.0\pm4.0$   &   \citeNP{naidu17}  \\
       &                &                  &                 &                           &   $2.9$          &   \citeNP{basu17}   \\ 
       &                &                  &                 &                           &   $5.6\pm0.7$    &   --do--            \\
 105   &  B1907+03      &  J1910+0358      &   2.3303        &    $3.27 \times 10^{12}$   &   $4.0\pm0.2$    &   \citeNP{weisb86}  \\
 106   &  B1911-04      &  J1913-0440      &   0.8259        &    $1.85 \times 10^{12}$   &   $\le0.5$       &   \citeNP{ritch76}  \\
 107   &  J1916+1023    &  J1916+1023      &   1.6183        &    $1.06 \times 10^{12}$   &   $47.0\pm4.0$   &   \citeNP{wang07b}  \\
 108   &  B1917+00      &  J1919+0021      &   1.2723        &    $3.16 \times 10^{12}$   &   $\le0.4$       &   \citeNP{ranki86}  \\
 109   &  J1920+1040    &  J1920+1040      &   2.2158        &    $3.83 \times 10^{12}$   &   $50.0\pm4.0$   &   \citeNP{wang07b}  \\
 110   &  B1918+19      &  J1921+1948      &   0.8210        &    $8.68 \times 10^{11}$   &   $9.0$, $43.0$  &   \citeNP{ranki13}  \\
 111   &  B1919+21      &  J1921+2153      &   1.3373        &    $1.36 \times 10^{12}$   &   $\le0.25$      &   \citeNP{ritch76}  \\
 112   &  J1926-1314    &  J1926-1314      &   4.8643        &    $1.35 \times 10^{13}$   & $\sim75.7\pm1.9$ &   \citeNP{rosen13}  \\
 113   &  B1923+04      &  J1926+0431      &   1.0741        &    $1.64 \times 10^{12}$   &   $\le5.0$       &   \citeNP{weisb86}  \\
 114   &  B1929+10      &  J1932+1059      &   0.2265        &    $5.18 \times 10^{11}$   &   $\le1.0$       &   \citeNP{ritch76}  \\
 115   &  B1933+16      &  J1935+1616      &   0.3587        &    $1.48 \times 10^{12}$   &   $\le0.06$      &   \citeNP{biggs92}  \\
 116   &  B1942+17      &  J1944+1755      &   1.9969        &    $1.22 \times 10^{12}$   &   $\le60.0$      &   \citeNP{lorim02}  \\
 117   &  B1942-00      &  J1945-0040      &   1.0456        &    $7.57 \times 10^{11}$   &   $21.0\pm1.0$   &   \citeNP{weisb86}  \\
 118   &  B1944+17      &  J1946+1805      &   0.4406        &    $1.04 \times 10^{11}$   &   $55.0\pm5.0$   &   \citeNP{yang14}   \\
       &                &                  &                 &                           &   $64.0\pm32.0$  &   \citeNP{ritch76}  \\
 119   &  B1946+35      &  J1948+3540      &   0.7173        &    $2.28 \times 10^{12}$   &   $\le0.75$      &   \citeNP{ritch76}  \\
 120   &  B2003-08      &  J2006-0807      &   0.5809        &    $1.65 \times 10^{11}$   &   $15.5\pm1.0$   &   \citeNP{basu17}   \\
 121   &  B2016+28      &  J2018+2839      &   0.5580        &    $2.91 \times 10^{11}$   &   $1.0\pm3.0$    &   \citeNP{naidu17}  \\
 122   &  B2020+28      &  J2022+2854      &   0.3434        &    $8.16 \times 10^{11}$   &   $0.2\pm1.6$    &   \citeNP{gajja17a} \\
 123   &  B2021+51      &  J2022+5154      &   0.5292        &    $1.29 \times 10^{12}$   &   $1.4\pm0.7$    &   \citeNP{gajja17a} \\
 124   &  J2033+0042    &  J2033+0042      &   5.0134        &    $7.21 \times 10^{12}$   &   $44-49, 53-58$ &   \citeNP{lynch13}  \\
 125   &  B2034+19      &  J2037+1942      &   2.0744        &    $2.08 \times 10^{12}$   &   $44.0\pm4.0$   &   \citeNP{herfi09}  \\
       &                &                  &                 &                           &   $24.2\pm1.5$   &  --do--             \\
       &&&&&& \\ \hline
\end{tabular}
\end{minipage}
\end{table*}
\begin{table*} 
  \caption{Continuation of Table~\ref{t_list01}, \ref{t_list02} \& \ref{t_list03}.}
  \label{t_list04}
\begin{minipage}{0.95\textwidth}
\begin{tabular}{|r l l l l l l|} \hline
       &&&&&& \\
       &  PSR Name      &    J-Name        &   P$_{\rm s}$     &    B$_{\rm s}$              &   NF               & References        \\
       &                &                  &   (s)           &    (G)                    &   (\%)           &                   \\ 
       &&&&&& \\
 126   &  B2044+15      &  J2046+1540      &   1.1383        &    $4.61 \times 10^{11}$   &   $\le0.04$      &   \citeNP{weisb86}  \\
 127   &  B2045-16      &  J2048-1616      &   1.9616        &    $4.69 \times 10^{12}$   &   $22.0\pm5.0$   &   \citeNP{naidu17}  \\
       &                &                  &                 &                           &   $5.5\pm0.2$    &   \citeNP{basu18}   \\
 128   &  B2053+36      &  J2055+3630      &   0.2215        &    $2.89 \times 10^{11}$   &   $\le0.7$       &   \citeNP{weisb86}  \\
 129   &  B2110+27      &  J2113+2754      &   1.2028        &    $1.78 \times 10^{12}$   &   $\le30.0$      &   \citeNP{redma09}  \\
 130   &  B2111+46      &  J2113+4644      &   1.0147        &    $8.62 \times 10^{11}$   &   $21.0\pm4.0$   &   \citeNP{gajja17a} \\
 131   &  B2113+14      &  J2116+1414      &   0.4402        &    $3.61 \times 10^{11}$   &   $\le1.0$       &   \citeNP{weisb86}  \\
 132   &  B2122+13      &  J2124+1407      &   0.6941        &    $7.39 \times 10^{11}$   &   $\le22.0$      &   \citeNP{redma09}  \\
 133   &  B2154+40      &  J2157+4017      &   1.5253        &    $2.32 \times 10^{12}$   &   $7.5\pm2.5$    &   \citeNP{ritch76}  \\
 134   &  J2208+5500    &  J2208+5500      &   0.9332        &    $2.58 \times 10^{12}$   &   $\le7.5$       &   \citeNP{joshi09}  \\
 135   &  B2217+47      &  J2219+4754      &   0.5385        &    $1.23 \times 10^{12}$   &   $\le2.0$       &   \citeNP{ritch76}  \\
 136   &  J2253+1516    &  J2253+1516      &   0.7922        &    $2.32 \times 10^{11}$   &   $\le49.0$      &   \citeNP{redma09}  \\
 137   &  B2303+30      &  J2305+3100      &   1.5759        &    $2.16 \times 10^{12}$   &   $1.0$          &   \citeNP{ranki86}  \\
 138   &  B2310+42      &  J2313+4253      &   0.3494        &    $2.01 \times 10^{11}$   &   $\le11.0$      &   \citeNP{redma09}  \\
 139   &  B2315+21      &  J2317+2149      &   1.4447        &    $1.24 \times 10^{12}$   &   $3.0\pm0.5$    &   \citeNP{weisb86}  \\
 140   &  B2319+60      &  J2321+6024      &   2.2565        &    $4.03 \times 10^{12}$   &   $29.0\pm1.0$   &   \citeNP{gajja17a} \\
 141   &  B2327-20      &  J2330-2005      &   1.6436        &    $2.79 \times 10^{12}$   &   $12.0\pm1.0$   &   \citeNP{biggs92}  \\
 142   &  J2346-0609    &  J2346-0609      &   1.1815        &    $1.28 \times 10^{12}$   &   $42.5\pm3.8$   &   \citeNP{basu17}   \\
       &                &                  &                 &                           &   $28.7\pm1.8$   &   \citeNP{basu17}   \\
       &&&&&& \\ \hline
\end{tabular}
\end{minipage}
\end{table*}
\begin{table*} 
  \caption{P$_{\rm   s}$ and B$_{\rm  s}$ of known nulling pulsars for which no NF estimates are available.}
  \label{t_list05}
\begin{minipage}{0.95\textwidth}
\begin{tabular}{|r l l l l l|} \hline
       &&&&& \\
       &  PSR Name      &    J-Name        &   P$_{\rm s}$     &    B$_{\rm s}$              & References        \\
       &                &                  &   (s)           &    (G)                    &                   \\ 
       &&&&& \\
   1   &  J0229+20      &  J0229+20        &   0.8069        &    NA                     &   \citeNP{denev13}  \\
   2   &  J0726-2612    &  J0726-2612      &   3.4423        &    $3.21 \times 10^{13}$   &   \citeNP{burke12}  \\
   3   &  B0853-33      &  J0855-3331      &   1.2675        &    $2.86 \times 10^{12}$   &   \citeNP{burke12}  \\
   4   &  J0941-39      &  J0941-39        &   0.5868        &    NA                     &   \citeNP{burke10}  \\
   5   &  J0943+2253    &  J0943+2253      &   0.5330        &    $2.21 \times 10^{11}$   &   \citeNP{brink18}  \\
   6   &  J1012-5830    &  J1012-5830      &   2.1336        &    $9.07 \times 10^{12}$   &   \citeNP{burke12}  \\
   7   &  J1055-6905    &  J1055-6905      &   2.9193        &    $7.80 \times 10^{12}$   &   \citeNP{burke12}  \\
   8   &  B1056-57      &  J1059-5742      &   1.1850        &    $2.28 \times 10^{12}$   &   \citeNP{burke12}  \\
   9   &  J1129-53      &  J1129-53        &   1.0629        &    NA                     &   \citeNP{burke12}  \\
  10   &  B1131-62      &  J1133-6250      &   1.0229        &    $6.88 \times 10^{11}$   &   \citeNP{burke12}  \\
  11   &  B1154-62      &  J1157-6224      &   0.4005        &    $1.27 \times 10^{12}$   &   \citeNP{burke12}  \\ 
  12   &  J1225-6035    &  J1225-6035      &   0.6263        &    $4.30 \times 10^{11}$   &   \citeNP{burke12}  \\
  13   &  J1255-6131    &  J1255-6131      &   0.6580        &    $1.64 \times 10^{12}$   &   \citeNP{burke12}  \\
  14   &  J1307-6318    &  J1307-6318      &   4.9624        &    $1.04 \times 10^{13}$   &   \citeNP{burke12}  \\
  15   &  B1323-58      &  J1326-5859      &   0.4780        &    $1.26 \times 10^{12}$   &   \citeNP{burke12}  \\
  16   &  B1323-63      &  J1326-6408      &   0.7927        &    $1.59 \times 10^{12}$   &   \citeNP{burke12}  \\
  17   &  J1406-5806    &  J1406-5806      &   0.2883        &    $4.25 \times 10^{11}$   &   \citeNP{burke12}  \\
  18   &  J1423-6953    &  J1423-6953      &   0.3334        &    $7.04 \times 10^{11}$   &   \citeNP{burke12}  \\
  19   &  B1424-55      &  J1428-5530      &   0.5703        &    $1.10 \times 10^{12}$   &   \citeNP{burke12}  \\
  20   &  B1449-64      &  J1453-6413      &   0.1795        &    $7.10 \times 10^{11}$   &   \citeNP{burke12}  \\
  21   &  B1454-51      &  J1457-5122      &   1.7483        &    $3.08 \times 10^{12}$   &   \citeNP{burke12}  \\
  22   &  B1510-48      &  J1514-4834      &   0.4548        &    $6.56 \times 10^{11}$   &   \citeNP{burke12}  \\
  23   &  J1514-5925    &  J1514-5925      &   0.1488        &    $6.63 \times 10^{11}$   &   \citeNP{burke12}  \\
  24   &  B1555-55      &  J1559-5545      &   0.9572        &    $4.48 \times 10^{12}$   &   \citeNP{burke12}  \\
  25   &  J1624-4613    &  J1624-4613      &   0.8712        &    $2.33 \times 10^{11}$   &   \citeNP{burke12}  \\
  26   &  B1630-44      &  J1633-4453      &   0.4365        &    $1.67 \times 10^{12}$   &   \citeNP{burke12}  \\
  27   &  B1641-68      &  J1646-6831      &   1.7856        &    $1.76 \times 10^{12}$   &   \citeNP{burke12}  \\
  28   &  J1647-3607    &  J1647-3607      &   0.2123        &    $1.67 \times 10^{11}$   &   \citeNP{burke12}  \\
  29   &  J1649-4349    &  J1649-4349      &   0.8707        &    $1.98 \times 10^{11}$   &   \citeNP{burke12}  \\
  30   &  B1650-38      &  J1653-3838      &   0.3050        &    $9.33 \times 10^{11}$   &   \citeNP{burke12}  \\
  31   &  J1707-4729    &  J1707-4729      &   0.2665        &    $6.53 \times 10^{11}$   &   \citeNP{burke12}  \\
  32   &  J1736-2457    &  J1736-2457      &   2.6422        &    $3.05 \times 10^{12}$   &   \citeNP{burke12}  \\
  33   &  J1741-3016    &  J1741-3016      &   1.8938        &    $4.18 \times 10^{12}$   &   \citeNP{burke12}  \\
  34   &  J1742-4616    &  J1742-4616      &   0.4124        &    $1.19 \times 10^{11}$   &   \citeNP{burke12}  \\
  35   &  J1749+16      &  J1749+16        &   2.3117        &    NA                     &   \citeNP{denev16}  \\
  36   &  J1750+07      &  J1750+07        &   1.9088        &    NA                     &   \citeNP{denev16}  \\
  37   &  B1747-31      &  J1750-3157      &   0.9104        &    $4.28 \times 10^{11}$   &   \citeNP{burke12}  \\
  38   &  J1757-2223    &  J1757-2223      &   0.1853        &    $3.85 \times 10^{11}$   &   \citeNP{burke12}  \\
  39   &  J1758-2540    &  J1758-2540      &   2.1073        &    $1.83 \times 10^{12}$   &   \citeNP{burke12}  \\
  40   &  B1806-21      &  J1809-2109      &   0.7024        &    $1.66 \times 10^{12}$   &   \citeNP{burke12}  \\
  41   &  J1819-1458    &  J1819-1458      &   4.2632        &    $5.01 \times 10^{13}$   &   \citeNP{burke12}  \\
  42   &  J1823-1126    &  J1823-1126      &   1.8465        &    $8.31 \times 10^{12}$   &   \citeNP{burke12}  \\
  43   &  B1822-14      &  J1825-1446      &   0.2792        &    $2.55 \times 10^{12}$   &   \citeNP{burke12}  \\
  44   &  J1827-0750    &  J1827-0750      &   0.2705        &    $6.54 \times 10^{11}$   &   \citeNP{burke12}  \\
  45   &  J1830-1135    &  J1830-1135      &   6.2216        &    $1.74 \times 10^{13}$   &   \citeNP{burke12}  \\
       &&&&& \\ \hline
\end{tabular}
\end{minipage}
\end{table*}
\begin{table*} 
  \caption{Continuation of Table~\ref{t_list05}.}
  \label{t_list06}
\begin{minipage}{0.95\textwidth}
\begin{tabular}{|r l l l l l|} \hline
       &&&&& \\
       &  PSR Name      &    J-Name        &   P$_{\rm s}$     &    B$_{\rm s}$              & References        \\
       &                &                  &   (s)           &    (G)                    &                   \\ 
       &&&&& \\
  46   &  B1834-06      &  J1837-0653      &   1.9058        &    $1.23 \times 10^{12}$   &   \citeNP{burke12}  \\
  47   &  J1837-1243    &  J1837-1243      &   1.8760        &    $8.38 \times 10^{12}$   &   \citeNP{burke12}  \\
  48   &  J1840-1419    &  J1840-1419      &   6.5976        &    $6.54 \times 10^{12}$   &   \citeNP{burke12}  \\
  49   &  J1841-0310    &  J1841-0310      &   1.6577        &    $7.54 \times 10^{11}$   &   \citeNP{burke12}  \\
  50   &  J1852-0635    &  J1852-0635      &   0.5242        &    $2.78 \times 10^{12}$   &   \citeNP{burke12}  \\
  51   &  J1854-1557    &  J1854-1557      &   3.4532        &    $3.99 \times 10^{12}$   &   \citeNP{burke11}  \\
  52   &  J1857-1027    &  J1857-1027      &   3.6872        &    $6.31 \times 10^{12}$   &   \citeNP{burke12}  \\
  53   &  J1935+1159    &  J1935+1159      &   1.9398        &    $1.37 \times 10^{12}$   &   \citeNP{brink18}  \\
  54   &  B2043-04      &  J2046-0421      &   1.5469        &    $1.53 \times 10^{12}$   &   \citeNP{naidu17}  \\
  55   &  J2050+1259    &  J2050+1259      &   1.2210        &    $7.94 \times 10^{11}$   &   \citeNP{brink18}  \\
       &&&&& \\ \hline
\end{tabular}
\end{minipage}
\end{table*}
\begin{table*} 
  \caption{P$_{\rm   s}$ and B$_{\rm  s}$ of known Intermittent Pulsars.}
  \label{t_list07}
\begin{minipage}{0.95\textwidth}
\begin{tabular}{|r l l l l l|} \hline
       &&&&& \\
       &  PSR Name      &    J-Name        &   P$_{\rm s}$     &    B$_{\rm s}$              & References        \\
       &                &                  &   (s)           &    (G)                    &                   \\ 
       &&&&& \\
   1   &  J1107-5907    &  J1107-5907      &   0.2528        &    $4.83 \times 10^{10}$   & \citeNP{meyer18}  \\
   2   &  J1832+0029    &  J1832+0029      &   0.5339        &    $9.09 \times 10^{11}$   & \citeNP{lorim12}  \\
   3   &  J1839+15      &  J1839+15        &   0.5492        &    $3.83 \times 10^{12}$   & \citeNP{surni13}  \\
   4   &  J1841-0500    &  J1841-0500      &   0.9129        &    $5.80 \times 10^{12}$   & \citeNP{camil12}  \\
   5   &  J1910+0517    &  J1910+0517      &   0.3080        &    $4.80 \times 10^{11}$   & \citeNP{lyne17}   \\
   6   &  J1929+1357    &  J1929+1357      &   0.8669        &    $1.80 \times 10^{12}$   & \citeNP{lyne17}   \\
   7   &  B1931+24      &  J1933+2421      &   0.8137        &    $2.60 \times 10^{12}$   & \citeNP{krame06a} \\
       &&&&& \\ \hline
\end{tabular}
\end{minipage}
\end{table*}

\end{document}